\documentclass[fleqn,twoside]{article}
\usepackage{espcrc2}

\usepackage{epsfig}
\usepackage[figuresright]{rotating}

\newcommand{\AmS}{{\protect\the\textfont2
  A\kern-.1667em\lower.5ex\hbox{M}\kern-.125emS}}
\newcommand{\bes}{\begin{eqnarray}}
\newcommand{\ees}{\end{eqnarray}}
\newcommand{\bea}{\begin{array}}
\newcommand{\ea}{\end{array}}

\newcommand{\fig}[1]{Fig.~\ref{#1}}
\newcommand\norm[1]{\|#1\|}
\newcommand\Enorm[1]{\norm{#1}_{\rm E}}
\def\mdm{M^{\dagger}M}
\def\mdme{(\mdm)_e}
\newcommand{\ev}[1]{\langle #1 \rangle}
\def\rloc{r_{\rm loc}}
\def\rlocm{r_{\rm loc}^{\rm max}}
\def\mG{m_{\rm G}}

\title{
{
\vspace{-3.1cm} \normalsize \hfill
\parbox{30mm}{HU-EP-04/46\\SFB/CPP-04-34\\DESY 04-157\\August 2004}
}\\[15mm]
       The locality problem for two tastes of staggered fermions
       \thanks{Talk presented by F. Knechtli.}
       }

\author{B. Bunk\address[HU]{Institut f{\"u}r Physik, 
        Humboldt Universit{\"a}t, Newtonstr. 15, 12489 Berlin, Germany},
        M. Della Morte\addressmark[HU],
        K. Jansen\address[DNIC]{NIC/DESY-Zeuthen, Platanenallee 6, 15738 Zeuthen,
        Germany}
        and F. Knechtli\addressmark[HU]
       }

\begin{document}

\begin{abstract}
We address the locality problem arising in simulations, which take the
square root of the staggered fermion determinant as a Boltzmann weight
to reduce the number of dynamical quark tastes from four to two.
We study analytically and numerically the
square root of the staggered fermion operator as a candidate
to define a two taste theory from first principles.
Although it has the correct
weight, this operator is non-local in the continuum limit.
Our work serves as a warning that fundamental properties of field
theories
might be violated when employing blindly the square root trick.
The question, whether a local operator reproducing the square root of the
staggered fermion determinant exists, is left open.
\vspace{-0.5cm}
\end{abstract}

\maketitle

\section{INTRODUCTION}
\vspace{-0.1cm}
The naive discretization of the Dirac operator on a Euclidean space-time
lattice produces a theory with 16 Dirac fermion copies or tastes.
The question whether this multiplicity can be reduced by a factor four has
a positive answer. At finite lattice spacing the naive Dirac operator can be brought
to a four-block diagonal form by spin diagonalization. Each block is a local
operator, the staggered Dirac operator, whose determinant is the fourth root
of the determinant of the naive operator and describes four tastes in the continuum
limit. The question which arises at this point is whether one can construct the
``fourth root'' (in the above sense) of the staggered operator.

To this aim one should start from the taste basis for staggered fermions. In the
free theory four degenerate tastes can be identified in momentum space \cite{Weisz}
but are described by non-local fields in coordinate space and so this formulation
is not suitable for introducing gauge fields. In coordinate space there is a basis
of four taste-fields constructed from paths which combine (in a gauge covariant way)
staggered fields living in a $2^4$ hypercube
\cite{Klube}. Even in the free theory there are taste-changing interactions,
which vanish as $a^2$ ($a$ is the lattice spacing) in the continuum limit.
We would like to stress that the mixing terms of the hypercube fields are
generated by a term with the structure like the Wilson term for Wilson fermions,
which prevents the doubling of the four hypercube fields in the continuum limit.

There is no obvious way of reducing the number of staggered fermion tastes.
For the Monte Carlo simulation of two degenerate (or one) dynamical quarks
one resorts to the so called square (or fourth) root trick \cite{trick}.
The Boltzmann weight is defined by taking the square (or fourth) root of the
determinant of the staggered Dirac operator, and this can be simulated
efficiently. As concerns the observables, quark propagators are computed using
the four-taste staggered operator and the sources project onto the desired
valence taste components \cite{tom}. This construction seems to work in
the classical continuum limit and in partially quenched chiral perturbation
theory \cite{begol}. It is not based though on a first principle formulation
of the fermionic theory, so far no acceptable operator describing two (or one)
tastes has been found. Hence causality and the existence of a universal continuum
limit of this theory cannot be established.

One of the requirements for universality to hold is that the Dirac operator
is local \cite{hjl,ferenc}. This means for its kernel in coordinate space $G(x,y)$
\bes
 \norm{G(x,y)} & \le & C{\rm e}^{-\gamma\Enorm{x-y}/a}\,,\quad
 \gamma>0\,, \label{locality}
\ees
where $\norm{G(x,y)}$ is the operator norm, $\Enorm{x-y}$ is the Euclidean norm
and $C$, $\gamma$ are numbers independent of the gauge field. Only local gauge
paths are allowed in the definition of the Dirac operator.

In this work we present a locality test for a candidate operator to describe
two tastes, namely the most naive choice, which is to take the square root of
the staggered operator itself. This operator has a determinant, which is obviously
equal to the square root of the staggered determinant,
but turns out to be non-local.
The outcome might not be a surprise given the non-analyticity of the
square root and the blindness concerning the taste structure.
Nevertheless our investigation can be considered as a benchmark study
in the spirit of Ref. \cite{hjl} for a still to be found operator.
Here we only summarize our main results,
which are published in detail in Ref. \cite{local}.


\section{ANALYTIC STUDY OF A CANDIDATE TWO-TASTE OPERATOR}
\vspace{-0.1cm}
We denote by $M$ the staggered Dirac operator.
In numerical simulations the identity
\bes
 \det(aM) & = & \det(a^2\mdme) \label{detmo}
\ees
is useful, where $\mdme$ acts on fields defined on the even sites of the
lattice only. We investigate the operator
\bes
 D & = & \sqrt{(\mdm)_e} \,, \label{ourD}
\ees
where the square root is computed through a series of Chebyshev polynomials.

In the free theory
\bes
 \mdm & = & m^2-\Delta(2a)\,, \label{laplace}
\ees
where $m$ is the bare quark mass and $\Delta(2a)$ is the Laplace operator
on a lattice with spacing $2a$.
The kernel $D(x)$ can be computed analytically by Fourier transformation from
momentum space. The result in the continuum limit is
\bes
   \frac{-3}{4\pi^2\Enorm{x}^5} e^{-m\Enorm{x}}
   \left( 1 + m\Enorm{x} + \frac{m^2\Enorm{x}^2}{3} \right)
\ees
and means that the free operator $D$ is non-local at the scale
\bes
 \rloc^{({\rm free})} & = & 1/m \,.\label{rlocfree}
\ees
It is very unlikely that the interacting theory will turn out to be local.
At best one can hope that the localization range $\rloc$ will be proportional
to the inverse of a large hadron mass. This motivates our numerical study.
%
\begin{figure}[t]
\centerline{\epsfig{file=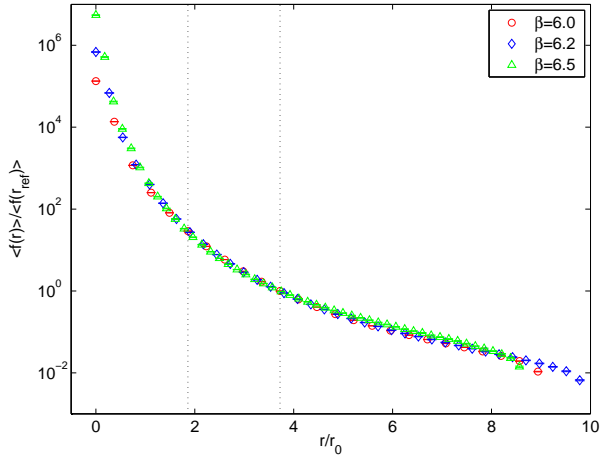,width=8cm}}
\vspace{-1.0cm}
\caption{Decay of $\ev{f(r)}/\ev{f(r_{\rm ref})}$, where $r_{\rm ref}$
is marked by the rightmost vertical dotted line. \label{f_f}} 
\vspace{-0.5cm}
\end{figure}
%
%
\begin{figure}[t]
\vspace{-0.15cm}
\centerline{\epsfig{file=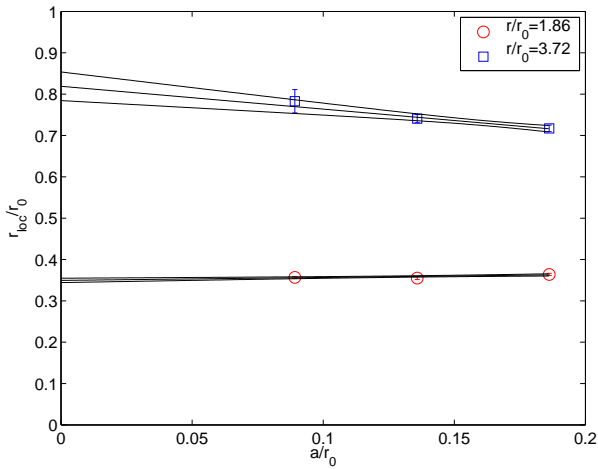,width=8cm}}
\vspace{-1.0cm}
\caption{Effective localization range from $\ev{f(r)}$ computed at the
two distances $r$ marked by the vertical dotted lines in \fig{f_f}.
\label{f_rloc}} 
\vspace{-0.5cm}
\end{figure}
%

\section{NUMERICAL RESULTS}
\vspace{-0.1cm}
We work in the quenched approximation of QCD and generate gauge configurations
at three values of the lattice spacing corresponding to $\beta=6.0,\;6.2,\;6.5$
with the Wilson gauge action. The bare quark mass $m$ is tuned to keep
the mass $\mG$ of the Goldstone pion $\pi_{\rm G}$
at the value $r_0\mG=1.30(3)$ ($r_0$ is the Sommer scale). We have two sets
of volumes with lattice sizes $L\mG\approx4$ and $L\mG\approx6$.
To study the locality of $D$ we follow closely Ref. \cite{hjl}.
On each configuration
we define a source field $\xi_c(x)$ to be 1 for one point $x=y$
(we always set $y=0$) and one color component $c=1$, and zero otherwise.
We compute the decay properties of $\psi(x)=aD\xi(x)$ in terms of
\bes
 f(r) & = & \max\big\{\norm{\psi(x)}\,\big|\,\norm{x-y}_1=r\big\} \,.
\ees
Given a distance $r$ from the source at $y$ we find the maximum of the SU(3)
color norm $\norm{\psi(x)}$ and denote it by $f(r)$. For the distance $r$
we take the ``taxi-driver distance''
\bes
 \norm{x-y}_1 = \sum_{\mu}\min\{|x_\mu-y_\mu|,L-|x_\mu-y_\mu|\}
\ees
and its largest possible value is $2L$.

In \fig{f_f} we plot on a logarithmic scale the gauge average
$\ev{f(r)}$ (normalized at the distance $r_{\rm ref}/r_0=3.72$) as a function
of the distance $r$ in units of $r_0$. The results shown are for the volume
$L\mG\approx6$ and the three lattice spacings. There is remarkable scaling as
the continuum limit is approached, indeed suggesting a physical localization
range. We assume the form
\bes
 \ev{f(r)} & \propto & {\rm e}^{-r/\rloc(r)}
\ees
and determine an inverse effective localization range $\rloc(r)^{-1}$ using two
consecutive values of $\ev{f(r)}$. The value of $\rloc(r)$ increases with
$r$ and has a ``bump'' at large distances $r$. This ``bump'' becomes higher
as the physical lattice size $L$ is increased and will eventually turn into a
plateau on very large volumes, where there are no more finite volume effects.
We understand the main origin of the latter \cite{local}.

We adopt the strategy to compute the localization range at a fixed physical value
of the distance $r$. For the choices $r/r_0=1.86$ and $r/r_0=3.72$ (marked by
the vertical dotted lines in \fig{f_f}) we checked that there are no finite
volume effects in $\rloc(r)$ computed in the volume $L\mG\approx6$. The results
for the different lattice spacings are shown in \fig{f_rloc} and the continuum
extrapolation yields
\bes
 \rloc(r/r_0=1.86)\mG & = & 0.455(12) \,,\\
 \rloc(r/r_0=3.72)\mG & = & 1.06(5) \,.
\ees
A local operator has $\rloc(r)={\rm O}(a)$ for all distances $r$.
The largest value of $\rloc(r)$ in the volume $L\mG\approx6$ has the
continuum limit
\bes
 \rlocm\mG & = & 2.8(6) \,.
\ees
So we conclude that in the interacting case the scale of the inverse localization
range is most likely given by the lightest pion mass.

Our present work leaves the question open, whether a local operator $D$ exists
whose determinant is equal to the square root of the determinant of the
staggered Dirac operator.
We emphasize that present dynamical simulations, where the Boltzmann
weight is taken to be the square root of the staggered determinant, do not use the
square root of the staggered operator.
If a local operator $D$ is found then the configurations generated in these
simulations are safe. There still remains however the unitarity problem. $D$ will
dictate the appropriate Green's functions for a two-taste theory and most likely
these are not the ones built with the four-taste staggered operator.

{\bf Acknowledgement.} We thank C. Bernard for a discussion during the
conference.
The simulations in the present work have been carried out using
the MILC collaboration's public lattice gauge theory code. We also
thank the computer center at DESY Zeuthen and Humboldt University for their
professional assistance. This work was supported in part by 
by the Deutsche Forschungsgemeinschaft in the SFB/TR 09.

\end{document}